\documentclass[twocolumn]{aastex631}
\usepackage{amsmath,amssymb}
\usepackage{acronym}
\newcommand{\chib}{\chi_{\mathrm{b}}}
\newcommand{\alphaCE}{\alpha_\mathrm{CE}}

\newcommand{\mchirp}{\mathcal{M}}
\usepackage{printlen}
\usepackage{booktabs}

\DeclareMathOperator{\logit}{logit}
\DeclareMathOperator{\arctanh}{arctanh}

\newcommand{\changed}[1]{{#1}}

\begin{document}
\input{result_macros_extra.sty}

\acrodef{GW}[GW]{gravitational wave}
\acrodef{BBH}[BBH]{binary black hole}
\acrodef{AMAZE}[AMA$\mathcal{Z}$E]
{Astrophysical Model Analysis and Evidence Evaluation}
\acrodef{DNN}[DNN]{deep neural network}
\acrodef{GPR}[GPR]{Gaussian process regression}
\acrodef{KDE}[KDE]{kernel density estimation}

\acrodef{CIERA}[CIERA]{Center for Interdisciplinary Exploration and Research in Astrophysics}
\acrodef{STFC}[STFC]{Science and Technology Facilities Council}

\title{Exploring the evolution of gravitational-wave emitters with efficient emulation: Constraining the origins of binary black holes using normalising flows}
\shorttitle{Constraining the origins of binary black holes using normalising flows}

\author[0009-0009-9828-3646]{Storm Colloms}
\affiliation{Institute for Gravitational Research, University of Glasgow, Kelvin Building, University Avenue, Glasgow, G12 8QQ, Scotland}
\correspondingauthor{Storm Colloms}
\email{s.colloms.1@research.gla.ac.uk}

\author[0000-0003-3870-7215]{Christopher P L Berry}
\affiliation{Institute for Gravitational Research, University of Glasgow, Kelvin Building, University Avenue, Glasgow, G12 8QQ, Scotland}
\affiliation{Center for Interdisciplinary Exploration and Research in Astrophysics (CIERA), Northwestern University, 1800 Sherman Avenue, Evanston, IL 60201, USA}

\author[0000-0002-6508-0713]{John Veitch}
\affiliation{Institute for Gravitational Research, University of Glasgow, Kelvin Building, University Avenue, Glasgow, G12 8QQ, Scotland}

\author[0000-0002-0147-0835]{Michael Zevin}
\affiliation{The Adler Planetarium, 1300 South DuSable Lake Shore Drive, Chicago, IL 60605, USA}
\affiliation{Center for Interdisciplinary Exploration and Research in Astrophysics (CIERA), Northwestern University, 1800 Sherman Avenue, Evanston, IL 60201, USA}
\affiliation{NSF-Simons AI Institute for the Sky (SkAI), 172 E.\ Chestnut Street, Chicago, IL 60611, USA}
 
\begin{abstract}

Binary population synthesis simulations allow detailed modelling of gravitational-wave sources from a variety of formation channels. 
These population models can be compared to the observed catalogue of merging binaries to infer the uncertain astrophysical input parameters describing binary formation and evolution, as well as the relative rates between various formation pathways. 
However, it is computationally infeasible to run population synthesis simulations for all variations of uncertain input physics. 
We demonstrate the use of normalising flows to emulate population synthesis results and interpolate between astrophysical input parameters. 
Using current gravitational-wave observations of binary black holes, we use our trained normalising flows to infer branching ratios between multiple formation channels, and simultaneously infer common-envelope efficiency and natal spins across a continuous parameter range. 
Given our set of formation channel models, we infer the natal spin to be $\chibcontall$, and the common-envelope efficiency to be $>\alphaCEcontlowlim$ at $90\%$ credibility, with the majority of underlying mergers coming from the common-envelope channel.
Our framework allows us to measure population synthesis inputs where we do not have simulations, and better constrain the astrophysics underlying current gravitational-wave populations.

\end{abstract}

\keywords{Stellar mass black holes (1611); Gravitational wave astronomy (675); Gravitational wave sources (677); Multiple star evolution (2153); Bayesian statistics (1900); Astrostatistics techniques (1886)}

\section{Introduction} \label{sec:intro}

\Acp{GW} provide a unique insight into the properties of coalescing compact-object binaries. 
The first three LIGO--Virgo--KAGRA observing runs yielded $90$ probable \ac{GW} candidates \citep{LIGOScientific:2021usb, LIGOScientific:2021djp}, with hundreds more events expected to be detected over the ongoing fourth observing run \citep{KAGRA:2013rdx}.
Analysis of the population properties of these events gives insight into how merging \acp{BBH} can form and how their progenitors evolve \citep{Mapelli:2021taw, theligoscientificcollaborationPopulationMergingCompact2022, Callister:2024cdx}.
The sub-population of \acp{BBH} formed through the evolution of field binaries can inform us about  binary stellar evolution \citep{Wysocki:2017isg, giacobboMergingBlackHole2018, baveraImpactMasstransferPhysics2021, delfaveroIterativelyComparingGravitationalWave2023, godfreyCosmicCousinsIdentification2023}; whereas the sub-population formed from dynamical interactions can inform us about the properties of their dense cluster environments \citep{Romero-Shaw:2020siz, fishbachGlobularClusterFormation2023, Ng:2023wbx, Kritos:2025bby}.
The specifics of these formation processes are still poorly understood, such as the star formation rates at given metallicities, the details of binary stellar evolution and interactions, and the relative rates of \ac{BBH} mergers from different formation channels \citep{Belczynski:2021zaz}.
The uncertainty in model predictions is currently large, and differs depending on the initial input physics \citep{Mandel:2021smh, Belczynski:2021zaz, mandelMergingStellarmassBinary2022}. 
Population analysis of \ac{GW} detections offers a unique way to constrain the uncertainty from model predictions.

Hierarchical inference is used to infer properties of population models from the measured \ac{GW} observables, taking the measurement uncertainties and selection effects into account \citep{mandelExtractingDistributionParameters2019, thraneIntroductionBayesianInference2019, Vitale:2020aaz}.
Modelling of compact binary merger populations has followed several paths.
One method uses parametric models with a simple functional form (e.g., a power-law), to fit the observed distributions of masses, spins and merger redshifts, as well as correlations between these parameters \citep[e.g.,][]{fishbachWhereAreLIGO2017, talbotMeasuringBinaryBlack2018, Kimball:2020qyd, callisterWhoOrderedThat2021, theligoscientificcollaborationPopulationMergingCompact2022}.
The functional forms may be motivated by astrophysical theory or convenience.
This approach risks forcing population features not present in the data, or missing features which are not allowed by the model.
On the other hand, data-driven models do not assume any predefined shape of the population \citep[e.g.,][]{Mandel:2016prl,Rinaldi:2021bhm, tiwariVAMANAModelingBinary2021, golombSearchingStructureBinary2022, Edelman:2022ydv, Ray:2023upk, Heinzel:2023hlb, callisterParameterFreeTourBinary}.
They can therefore identify features in the data that are not captured by parametric models, as long as statistical fluctuations are accounted for. 
Both of these approaches do not directly use models of \ac{BBH} progenitor evolution, making it difficult to link observed population features to astrophysical theory.

We aim to gain more insight into \ac{BBH} astrophysics by using population synthesis simulations to predicted the observed population.
These simulations model the evolution of \ac{BBH} progenitors, predicting the resulting observable parameters of sources for different input astrophysics and formation scenarios, which will we refer to as hyperparameters.
Works comparing population synthesis to \ac{GW} observations have constrained sets of these  hyperparameters \citep{Stevenson:2017tfq, Zevin:2017evb, neijsselEffectMetallicityspecificStar2019, Bouffanais:2020qds, zevinOneChannelRule2021, baveraImpactMasstransferPhysics2021, Mastrogiovanni:2022ykr, stevensonConstraintsContributionsObserved2022}.
However, simulating just a single population requires many simulated binaries, and investigating varying input physics requires  simulating many populations.
Current population synthesis codes take $\mathcal{O}(10^2)$ CPU hours for a single population of $10^6$ isolated binaries \citep{andrews2024posydonversion2population}.
Therefore, a detailed exploration of uncertain physical processes that affect predicted \ac{BBH} populations makes using population synthesis much more computationally costly than using parametric or data-driven models.

Exploring more population hyperparameters in an analysis gives us more insight into the degeneracies and correlations between different model populations \citep{barrettAccuracyInferencePhysics2018, stevensonConstraintsContributionsObserved2022}.
Furthermore, there is evidence for multiple formation channels present in the \ac{GW} data \citep{zevinOneChannelRule2021, Cheng:2023ddt, Li:2023yyt}; each will have its own contribution to the total population. 
Joint inference of these parameters gives better constraints on the inferred population, and is necessary to avoid biased results coming from a neglected channel or variable that truly contributes to the astrophysical observations \citep{zevinOneChannelRule2021,Cheng:2023ddt}.
To understand the correlations between different input astrophysical hyperparameters, we require a comprehensive set of simulations of \ac{BBH} populations considering a broad range of simulation inputs.

Machine learning is a potential avenue for reducing the difficulties associated with these expensive simulations.
Machine learning has been used in various aspects of \ac{GW} data analysis in order to increase efficiency of data-quality improvement, waveform modelling, searches, and parameter estimation \citep{Cuoco:2024cdk}. 
In the context of \ac{BBH} populations, we can employ machine learning to reduce the cost of performing population synthesis simulations: instead of performing numerous simulations across the input hyperparameter space, we may interpolate between simulations. 
The interpolation may either be done by reconstructing the probability distribution across different input hyperparameters given a set of observations \citep{delfaveroIterativelyComparingGravitationalWave2023} or by emulating how the astrophysical population varies across hyperparameter space. 
Different approaches have previously been explored for population emulation. 
Early studies used \acl{GPR} \citep[\acsu{GPR};][]{Barrett:2016edh,taylorMiningGravitationalwaveCatalogs2018}. 
While this emulation was effective on a small number of dimensions, \ac{GPR} was shown to be limited at handling higher dimensional data with sufficient computational efficiency, and the principal component analysis needed to reduce the input dimensionality to \ac{GPR} gives an inevitable loss of information in the training data \citep{wongGravitationalWavePopulation2020, cheungTestingRobustnessSimulationbased2022}. 
\Acp{DNN} have been used to emulate populations, training the \acp{DNN} on \ac{KDE} reconstructions of the population \citep{mouldDeepLearningBayesian2022} or binned distributions \citep{Riley:2023jep}.
The reliance on the evaluation of the population distributions once by the \acp{KDE} and then again by the \acp{DNN} adds uncertainty to the result, while using bins requires a trade-off between having sufficient training data per bin and having sufficient bins to resolve all the features in the population. 
Accurately emulating population synthesis models across many hyperparameters is a challenging problem.

Normalising flows are a type of generative machine learning algorithm which can model complicated probability densities; they are therefore well suited to this problem. 
Normalising flows operate by learning a series of transformations from a simple distribution to a target distribution \citep{papamakariosNormalizingFlowsProbabilistic2021, kobyzevNormalizingFlowsIntroduction2021}. 
Applications of normalising flows to population inference problems have given fast and accurate results for population analyses of stellar mass \acp{BBH} \citep{wongGravitationalWavePopulation2020, Wong:2020yig, Wong:2020ise}, supermassive \acp{BBH} \citep{Laal:2024trp}, and studies of other stellar populations \citep{hon2024flowbasedgenerativeemulationgrids}.
\cite{wongGravitationalWavePopulation2020} demonstrated that normalising flows are capable of recovering astrophysical hyperparameters of simulated observations across a higher dimensional set of binary parameters than \ac{GPR}.
The application of normalising flows was extended by \cite{Wong:2020ise}, inferring two astrophysical hyperparameters and the branching ratio between two formation channels with GWTC-2 observations \citep[up to the first part of the third LIGO--Virgo--KAGRA observing run;][]{LIGOScientific:2020ibl}.
These studies show the effectiveness of normalising flows when applied to problems with increased dimensionality, using the latest sets of \ac{GW} observations.
Using normalising flows as emulators of population synthesis models allows for astrophysically informed inference of \ac{BBH} populations, while reducing the computational expense of simulating many populations.

We use normalising flows to emulate and interpolate population synthesis models conditional on the simulation input parameters, using this as our population model for hierarchical inference of two astrophysical hyperparameters and branching fractions between five different formation channels. 
We extend the \ac{AMAZE} framework used in \cite{zevinOneChannelRule2021} and \cite{Cheng:2023ddt} to incorporate the normalising flows and move to a continuous hyperparameter space of our population models.
The interpolation allows for the inference of the astrophysical hyperparameters between the originally simulated models, giving us insight into the uncertainties and correlations of these hyperparameters. 
We illustrate our approach using GWTC-3 observations \citep[up to the end of the third LIGO--Virgo--KAGRA observing run;][]{LIGOScientific:2021djp}.
In showing that our inference can reliably infer a continuous range of astrophysical hyperparameters, we open up possibilities of increasing the dimensionality of astrophysical hyperparameters studied in \ac{BBH} population inference.
Applying this method in cases of higher dimensional problems, together with our growing set of \ac{GW} observations would give key insights into in the formation of \acp{BBH} and their formation environments.

In Section~\ref{sec:methods} we underline the methods for the analysis including the population models, inference framework, normalising flows, and the \ac{GW} observations used. 
\changed{Further details on the training of the normalising flows are given in Appendix~\ref{apx:trainingflows}.}
We show results with our trained normalising flows in Section~\ref{sec:results}; these include tests of the normalising flows' emulation, inference comparison to a non-interpolating method, continuous inference with interpolation, and the resulting fit to the observed data.
In Section~\ref{sec:conc} we discuss the results and limitations of the current analysis, and avenues for future work. 
\changed{An accompanying data release is available from Zenodo \citep{amazeflowsDR}.}

\section{Methods}\label{sec:methods}

We infer astrophysical hyperparameters of the population of \acp{BBH} observed with \acp{GW}.
To do this, we use hierarchical Bayesian inference to measure the population hyperparameters given our observed GW population.
Our statistical framework is described in Section~\ref{met:statframe}.
This requires model \ac{BBH} populations from population synthesis, on which we train normalising flows, and then compare to the observations, which are each described in Section~\ref{met:popmodels}, Section~\ref{met:nflows}, and Section~\ref{met:obs}, respectively.
We update the \ac{AMAZE} framework developed by \citet{zevinOneChannelRule2021} to incorporate the addition of normalising flows.
Using normalising flows to interpolate between the population models, we evaluate the model likelihood across population hyperparameter space \emph{continuously}, therefore performing continuous inference of our astrophysical hyperparameters of interest.

\subsection{Population Models} \label{met:popmodels}

We consider five formation channels: three isolated channels (common envelope, chemically homogeneous evolution, stable mass transfer); and two dynamical cluster environments (globular clusters and nuclear star clusters).
These are the same population models as used in \citet{zevinOneChannelRule2021}, which contains further details of the simulations.

The common-envelope (CE) channel describes stellar binaries that, after one compact object has already formed, undergo unstable mass transfer leading to an envelope of the donor star's material surrounding the compact object and the donor's core. 
Drag forces from the envelope cause the system to harden, and the envelope is ejected with some efficiency, $\alphaCE$ \citep{vandenHeuvel1976, 1976Paczynski, Belczynski:2001uc, dominik2012}.
This efficiency is the fraction of orbital energy used to eject the envelope, where a value of $\alphaCE$ greater than $1$ requires energy other than orbital energy, such as radiative energy from the stellar objects, to be used in the envelope's ejection \citep{Ivanova:2012vx}.
A lower CE efficiency creates post-CE systems that are tighter, although at a risk of some systems merging in the envelope, while a higher efficiency creates systems that may still be too wide to merge within a Hubble time.
For the stable mass transfer (SMT) channel, binaries undergo only mass transfer which is stable during their evolution \citep{vdHeuvel2017, neijsselEffectMetallicityspecificStar2019, gallegos-garciaBinaryBlackHole2021, vanSon:2021zpk, brielUnderstandingHighmassBinary2022}.
The CE and SMT channels were modelled with and early version of the \texttt{POSYDON} framework \citep{fragosPOSYDONGeneralPurposePopulation2022, andrews2024posydonversion2population}: \texttt{COSMIC} population synthesis \citep{breivikCOSMICVarianceBinary2020} up until the end of the second mass-transfer episode, and the detailed stellar evolution code \texttt{MESA} \citep{Paxton2011, Paxton2013, Paxton2015, Paxton2018, Paxton2019} from the formation of a black hole--helium star binary until the formation of the second black hole. 

The chemically homogeneous evolution (CHE) case accounts for \acp{BBH} that form from stars which are tidally synchronized during core hydrogen burning, causing the stars to rapidly spin and their interiors to become chemically homogeneous, such that they do not expand during the red-giant phase \citep{deMink:2008iu, Mandel:2015qlu, dubuisson2020}.
The CHE channel models were adapted from \cite{dubuisson2020}, who used a \texttt{MESA} grid of binary simulations.

In the globular cluster (GC) and nuclear star cluster (NSC) channels, binaries undergo dynamical scatterings and captures due to their dense formation environments \citep{Heggie:1975tg, Fitchett1983, antonini2016}.
NSCs are generally more massive than GCs, thus their higher escape velocities mean that they can retain more black hole merger remnants, and have a significant fraction of hierarchical mergers.
The GC models used the cluster Monte Carlo code \texttt{CMC} from \cite{Rodriguez2019}, 
using the Binary Stellar Evolution (\texttt{BSE}) package of \cite{Hurley:2000pk, hurleyEvolutionBinaryStars2002}, with some updated evolutionary prescriptions from \texttt{COSMIC} \citep{breivikCOSMICVarianceBinary2020}.
The NSC channel models use the semi-analytical model from \cite{antoniniBlackHoleGrowth2019} to determine the merger rate of \acp{BBH}. 
This model assumes a regulation of the energy generated by \acp{BBH} by the process of two-body relaxation in the system.

The population models include associated weights for each simulated \ac{BBH}.
These weights account for the probability of binary formation at particular metallicities and the increasing co-moving volume probed at higher redshift. 
The populations are simulated over a range of different metallicity values.
For all population models other than GCs \citep[][appendix~A.1.3]{zevinOneChannelRule2021}, the distribution of metallicities at a given redshift follows a truncated log normal distribution around the empirical mean metallicity from \cite{Madau:2016jbv}. 
The associated metallicity weight of each sample is the density of the metallicity distribution given the birth redshift of each simulated binary.
The total weight is the factor of the co-moving volume probed at the redshift of the merger multiplied by the metallicity weight \citep{zevinOneChannelRule2021}.

Each simulation predicts the set binary observables $\Vec{\theta}$: the chirp mass $\mathcal{M}$, mass ratio $q=m_2/m_1$ (where $m_2$ is the less massive black hole component mass and $m_1$ is the more massive black hole component mass, such that $0\leq q\leq1$), effective inspiral spin $\chi_{\mathrm{eff}}$, and redshift $z$ at which the \ac{BBH} merges. 

Within each channel, two astrophysical hyperparameters, $\Vec{\lambda}$, are varied: the birth spin of black holes born in isolation (i.e., born without prior spin-up of the stellar core due to tidal interactions) $\chib$, and the CE efficiency $\alphaCE$. 
The value of $\alphaCE$ only influences the evolution of binaries from the CE channel. 

\changed{The birth spin of black holes $\chib$ is a common hyperparameter between all formation channels. 
For the field formation channels, black holes are given a minimum spin of $\chib$ upon formation. 
A field channel black hole forms with $\chi>\chib$ if tidal effects or mass transfer sufficiently spin up the black holes. 
For black holes in the dynamical models, all first-generation black holes formed from stellar collapse have a spin of $\chib$, while merger products are born with a spin of $\chi\sim0.7$ appropriate for nearly equal-mass mergers \citep{Pretorius:2005gq, Gonzalez:2006md, Buonanno:2007sv}. 
Varying the black hole birth spin is a proxy for the efficiency of angular momentum transfer in massive stars \citep{Belczynski:2017gds, Bavera:2020inc} 
It is expected that efficient angular momentum transfer results in low birth spins of black holes, giving a narrow range of $\chib\lesssim0.01$ \citep{Fuller:2019sxi}. 
The models with larger $\chib$ extend the assumption of a single birth spin value (when ignoring the effects of tidal spin-up or spin-up by mass transfer) to a case where there is some process during stellar evolution or collapse (potentially related to inefficient angular momentum transfer) that imparts a characteristic spin on newly formed black holes \citep{zevinOneChannelRule2021}. 
}

The population synthesis simulations were generated at fixed values of $\Vec{\lambda}$, $\chib^{\mathrm{*}} \in \{0,0.1,0.2,0.5\}$ and $\alphaCE^{\mathrm{*}}\in \{0.2,0.5,1,2,5\}$ \citep{zevinOneChannelRule2021}. 
These were the only points in parameter space considered in previous inferences, and we will use them as the training set for the normalising flows. 
Our inference parameter space is bound by the maximum and minimum values of the training models: for the continuous inference we explore $\chib$ over the range $[0,0.5]$ and $\alphaCE$ over the range $[0.2,5]$. 

We infer the branching ratios of the five formation channels $\Vec{\beta}=\{\beta_{\mathrm{CE}}, \beta_{\mathrm{CHE}}, \beta_{\mathrm{GC}}, \beta_{\mathrm{NSC}}, \beta_{\mathrm{SMT}}\}$ as a mixture model between the channels.
Therefore in total we determine $7$ population hyperparameters ($6$ independent since $\Vec{\beta}$ must be normalised) made up of $\Vec{\Lambda}=\{\Vec{\lambda}, \vec{\beta}\}$.

\subsection{Statistical Framework} 
\label{met:statframe}

We infer a continuous posterior distribution over hyperparameters $\vec{\lambda}$ in addition to the branching fractions $\vec{\beta}$, following a hierarchical inference framework \citep{mandelExtractingDistributionParameters2019, thraneIntroductionBayesianInference2019, Vitale:2020aaz}.
To do this, we construct a hyperlikelihood for the population hyperparameters, using the posterior samples on the \ac{GW} observables $\vec{\theta}$ from parameter estimation released in GWTC-2.1 and GWTC-3 \citep{ LIGOScientific:2021djp, LIGOScientific:2021usb}. 
This gives the hyperlikelihood as
\begin{equation}
p(\mathbf{d} \mid \vec{\Lambda}) \propto \prod_{i=1}^{N_{\mathrm {obs }}} \frac{1}{S_i \tilde{\xi}(\vec{\lambda})} \sum_j \beta_j \sum_{k=1}^{S_i} \frac{p\left(\vec{\theta}_i^k \middle| \vec{\lambda} \right)}{\pi\left(\vec{\theta}_i^k\middle| \varnothing \right)},
\end{equation}
where $\mathbf{d}$ is the set of observed GW data, $S_i$ is total number of posterior samples for the $i$th GW for $N_{\mathrm {obs }}$ detections, $\tilde{\xi}(\vec{\lambda})=\sum_j \beta_j \int P_{\operatorname{det}}(\vec{\theta}) p(\vec{\theta} \mid \vec{\lambda}) \mathrm{d} \vec{\theta}$ is the detection efficiency of each population model, $\pi(\vec{\theta}_i^k\mid \varnothing)$ is the prior on each GW source for each posterior sample in the LVK analysis. 
We are solely interested in the shape of the distributions of $\vec{\theta}$ and therefore implicitly marginalize over the expected number of detections \citep[e.g.,][]{Fishbach:2018edt}.

The hyperposterior for $\Vec{\Lambda}$ is sampled with \texttt{emcee} \citep{Foreman_Mackey_2013}, using $250$ walkers with $1000$ steps and a $40\%$ burn in.
We inspect the sampling chains to check the chain convergence, and check the autocorrelation length and sample until we reach at least $50$ times the autocorrelation length of the measured hyperparameters. 

The previous \ac{AMAZE} framework \citep{zevinOneChannelRule2021} evaluated $p(\vec{\theta}_i^k \mid \vec{\lambda})$ and $\tilde{\xi}(\vec{\lambda})$ at discrete $\vec{\lambda} = \{\chib^*, \alphaCE^*\}$, where there existed samples from the population synthesis simulations.
This work expands the inference to be over any values of $\vec{\lambda}$ within the prior bounds, using normalising flows to evaluate $p(\vec{\theta}_i^k \mid \vec{\lambda})$. 
The detection efficiency, $\tilde{\xi}(\vec{\lambda})$ is evaluated at each training population point by using a signal-to-noise ratio cut on the population samples to find $P_{\operatorname{det}}$, then calculated across $\vec{\lambda}$ using a piecewise cubic Hermite interpolating polynomial interpolation \citep{PCHIP_1984}.
This ensures a continuously differentiable detection efficiency across the populations in each channel, preserving monotonicity between the known detection efficiency points.
This assumes that the values of the detection efficiency we calculate from the population synthesis are representative of the whole range of populations we explore.
The interpolation of the detection efficiency allows the continuous inference framework to work together to calculate the hyperposterior for $\vec{\Lambda}$, using the interpolated population models by the normalising flows.

We assume an uninformative prior across $\Vec{\beta}$, given by a Dirichlet distribution with equal concentration parameters of $1$, such that ($0<\beta_{\mathrm{i}} <1$) for all $\beta_{\mathrm{i}}$ and $\sum_i \beta_{\mathrm{i}} = 1$.
We use a uniform prior on $\chib$ and log-uniform in $\alphaCE$, within the bounds of the training set.

As we have a finite number of samples for each population model, the tails of the population model distributions $p(\vec{\theta} \mid \vec{\lambda})$ are extrapolated from a small number of samples.
Evaluating the probability of the population models in the tails of the distribution therefore does not give a reliable estimate of the true probability of \ac{BBH} formation in these regions. 
While the absolute error in reconstructing the distributions may be small, the relative error needed to compare different channels can be significant. 
To alleviate this problem, we add a regularisation term to the probability determined by the population models, such that a uniform value determined by constant $N$ is added to $p(\vec{\theta} \mid \vec{\lambda})$, giving a regularised distribution
\begin{equation}
    \label{eq:regularisation}
    p(\vec{\theta}\mid\vec{\lambda})=\frac{N}{N+1}p_{\mathrm{pop}}(\vec{\theta}\mid\vec{\lambda})+\frac{1}{N+1}U(\vec{\theta}).
\end{equation}
Here, $U(\vec{\theta})$ is a uniform distribution over the population parameters, and $p_{\mathrm{pop}}(\vec{\theta}\mid\vec{\lambda})$ is the reconstructed population distribution for a channel. 
The additional regularisation assumes that $p(\vec{\theta} \mid \vec{\lambda})$ tends towards a uniform distribution in parts of the parameter space where there is not enough information from the population synthesis simulations.
This form would result from treating reconstruction of the distribution as an inference problem and applying Laplace's rule of succession \citep[][chapter~18]{Jaynes_2003}. 
We start with a uniform prior where we have no samples to inform our result, and as we increase the number of population synthesis samples $N$, we place increasing weight in the population model probability $p_{\mathrm{pop}}(\vec{\theta}\mid\vec{\lambda})$.
In parts of $\vec{\theta}$ parameter space where there are many samples, the probability $p(\vec{\theta} \mid \vec{\lambda})$ is approximately equal to the population model probability $p_{\mathrm{pop}}(\vec{\theta} \mid \vec{\lambda})$, but in the tails of the distribution, we tend towards an \changed{uninformative} uniform distribution.
Our analysis uses the same $N$ for each channel, chosen to be the smallest number of training samples out of all of the training sub-populations. 
This means there is no preference between channels in regions of $\vec{\theta}$ parameter space where we have no samples for any channel.

The inference mirrors the \ac{AMAZE} framework used in \cite{zevinOneChannelRule2021} and \cite{Cheng:2023ddt}.
This included a change to the calculation of the prior on each GW posterior sample from the work in \cite{Cheng:2023ddt}, which ensures that there were no prior values with extremely small support for any of the event samples used. 

\subsection{Normalising flows} 
\label{met:nflows}

Normalising flows are a class of generative neural networks that are trained to find the optimal mapping between a complicated distribution and a simple latent space through a series of transformations, allowing emulation of the more complex distribution~\citep{papamakariosNormalizingFlowsProbabilistic2021,kobyzevNormalizingFlowsIntroduction2021}.
We use normalising flows to represent the distributions of GW observables for each formation channel given the population hyperparameters $p_{\mathrm{pop}}(\vec{\theta}\mid\vec{\lambda})$.
As we have samples from these distributions, but not the probabilities over $\vec{\theta}$-space, normalising flows give us a way to directly evaluate these probability distributions. 

During training, a normalising flow learns a set of transformations that can be used to reconstruct the distribution of interest from the latent space.
For a single transformation $f$, where the target distribution $x$ relates to the latent distribution $\mathcal{Z}$ as $\mathcal{Z}=f(x)$, the relationship between the target and latent distributions is
\begin{equation}
p_{x}(x\mid\phi)=p_\mathcal{Z}\left(f(x\mid\phi)\right)\left|\frac{\partial f(x\mid\phi)}{\partial x}\right|,
\end{equation}
allowing the normalising flow to map from the target distribution $p_x(x|\phi)$ to the latent distribution $p_\mathcal{Z}(f(x|\phi))$, given some conditional parameter $\phi$. 
In our case, $x$ will be related to our source $\vec{\theta}$ parameters (Appendix~\ref{apx:trainingflows}) and $\phi$ will be our population hyperparameters $\vec{\lambda}$ for each channel.  
For a normalising flow the Jacobian of the transformation $|{\partial f(x)}/{\partial x}|$ is tractable, and the function $f$ is invertible. 
Samples are taken from this latent distribution, which has an analytic form and can be easily sampled from, before being transformed back through the inverse of the learned transformations for emulation of the target distributions.
Our transformations use a multidimensional Gaussian for the latent distribution.

The optimised series of transformations are found by minimising the loss function of the network. 
We use the Kullback--Leibler (KL) divergence \citep{KullbackLeibler1951} between the normalising flow's approximated target distribution $p_x^*(x\vert\phi)$ and the original target distribution $p_x(x\vert\phi)$.
As our training samples also have associated weights to account for the distribution of \acp{BBH} in cosmic time and metallicity, we train the normalising flows to learn the weighted distribution of these samples, $w(x\vert\phi)p_x(x\vert\phi)$, where $w(x\vert\phi)$ is the distribution of sample weights. 
The KL divergence between the weighted distribution and the emulated distribution produced by the flow in the target space can be written as 
\begin{widetext}
\begin{equation*}
\label{eq:weightedKL}
    \begin{split}
    D_{\mathrm{KL}} \left[w_x(x\vert\phi)p_x^*(x\vert\phi)|| p_x(x\vert\phi)\right] &=\int_{-\infty}^{+\infty} w_x(x\vert\phi)p_x^*(x\vert\phi) \ln \left(\frac{w_x(x\vert\phi)p_x^*(x\vert\phi)}{p_x(x\vert\phi)}\right) \mathrm{d} x  \\
    &=-\int_{-\infty}^{+\infty} w_x(x\vert\phi)p_x^*(x\vert\phi) \ln p_x(x\vert\phi)  \mathrm{d} x+\text {constant} \\
    &=  -\mathbb{E}_{p_x^*(x\vert\phi)}\left[w_x(x\vert\phi)\ln p_x(x\vert\phi)\right]+\text { constant} \\
    &=-\mathbb{E}_{p_x^*(x\vert\phi)}\left\{w_x(x\vert\phi) \left[\ln p_\mathcal{Z}\left(f(x\vert\phi)\right)+\ln \left|\operatorname{det} \frac{\partial f(x \vert\phi)}{\partial x}\right|\right]\right\}+\text { constant}.
    \end{split}
\end{equation*}
\end{widetext}
In training, we find the gradient of the KL divergence, and thus ignore the constant terms, ultimately minimising the loss, $\mathcal{L}$,
\begin{equation}
    \mathcal{L} =-\frac{1}{\mathcal{N}}\sum_{i=1}^{\mathcal{N}} w_x(x_i) \left[ \ln p_\mathcal{Z}\left(f(x_i)\right)+\ln \left|\operatorname{det} \frac{\partial f(x_i )}{\partial x}\right| \right],
\end{equation}
where $\mathcal{N}$ is the number of samples used in each batch of training.

When training the normalising flows, the training samples from populations synthesis are mapped with a set of logistic and hyperbolic transformations in order to improve the training and avoid hard boundaries in the parameter space of the GW observables. 
For original samples given by parameters $\vec{\theta} = \{\mathcal{M}_c, q, \chi_\mathrm{eff}, z\}$, the transformed parameters $\vec{\theta '}$ span the range from negative to positive infinity, with a similar spread of values across dimensions.
The specifics of these transformations are detailed in Appendix~\ref{apx:trainingflows}.

\subsection{Observational Data}
\label{met:obs}

We use results from GWTC-3 \citep{LIGOScientific:2021djp}, analysing candidates with a false alarm rate of less than $1~\mathrm{yr}^{-1}$, following previous population analyses with this catalogue \citep{theligoscientificcollaborationPopulationMergingCompact2022}. 
The parameter estimation samples used are the mixed samples from GWTC-2.1 and GWTC-3 \citep{LIGOScientific:2021djp, LIGOScientific:2021usb}, containing an equal number of samples from \texttt{IMRPhenomXPHM} \citep{Pratten:2020ceb} and \texttt{SEOBNRv4PHM} \citep{Ossokine:2020kjp} waveform approximants.
We additionally exclude events GW190521 \citep{LIGOScientific:2020iuh} and GW190814 \citep{LIGOScientific:2020zkf}, as these are outliers within the context of our population models.
The detection probabilities are calculated for the LIGO-Hanford, LIGO-Livingston, and Virgo network operating at \texttt{midhighlatelow} sensitivities, as a proxy for the sensitivity of the detectors in their third observing run \citep{KAGRA:2013rdx}.
These choices are consistent with the analyses of \cite{zevinOneChannelRule2021} and \cite{Cheng:2023ddt}, enabling us to compare our results.

As the code currently requires the same number of posterior samplers for each GW, we use 11,058 posterior samples per GW.
For GWs that had fewer than 11,058 posterior samples, we repeated the total set of posterior samples for that GW, and then chose to repeat a subset of the posterior samples, chosen randomly. 
Our choice of 11,058 total posterior samples minimised the number of samples in the repeated subset for all GWs, while allowing us to use more posterior samples from the GWs which had them.

\section{Results} \label{sec:results}

Here we evaluate the emulation performance of our trained normalising flows with a series of tests to determine that they are trained sufficiently well for the following continuous inference.
Our first tests look at the performance of the normalising flow in emulating the training data and interpolating across hyperparameter space (Section~\ref{results:emintp}).
We use the normalising flows for inference on a discrete hyperparameter space, and show that these results are in agreement with results using the previous  (non-interpolating) representation of the population models (Section~\ref{results:discinf}).
We then present the results from our extension to inference across a continuous hyperparameter space (Section~\ref{results:continf}).
Finally, we look at the inferred distribution of \ac{GW} observables, given the continuous inference results (Section~\ref{results:dataspace}).

\subsection{Emulation and Interpolation performance}
\label{results:emintp}

Before using the normalising flows for continuous inference, we test their capabilities of accurately capturing the training data.
We compare them to the training samples from the population synthesis models as well as to \ac{KDE} representations of the astrophysical models.
The KDE representations of the astrophysical models give a smooth probability density across $\vec{\theta}$ space.
They are only used to construct these distributions at the points in hyperparameter space where we have population synthesis results, i.e.,  $\{\chib^{\mathrm{*}},\alphaCE^{\mathrm{*}}\}$.
Our \ac{KDE} implementation is consistent with the \ac{KDE} model representations used in \cite{zevinOneChannelRule2021} and \cite{Cheng:2023ddt}.
Comparing the two methods enables us to examine the robustness of our results to the approach used to represent the populations. 

\begin{figure*}
    \centering
    \includegraphics[width=0.72\linewidth]{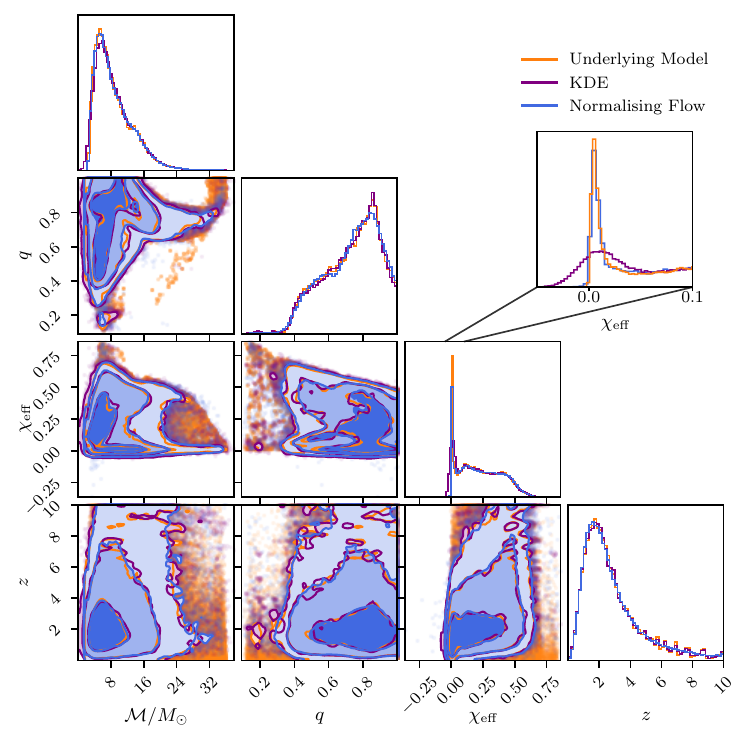}
    \caption{The distribution of GW observables from the training population synthesis samples (orange), the KDE of this population (purple), and the normalising flow (blue), for the CE channel with $\chib=0$ and $\alphaCE=2$.
    The confidence levels represent the areas containing $50\%$, $90\%$, and $99\%$ of the probability.
    The normalising flow matches the training distribution well both in the marginal distributions and in the correlations.
    The spike in the effective inspiral spin distribution, \changed{shown in more detail in the inset plot,} is due to the model assumption that black holes are born with a common spin.
    The number of effective samples is the same for all three model representations.
    }
    \label{fig:flowmodelcorner}
\end{figure*}

An example of the samples drawn from the trained normalising flows in $\vec{\theta}$-space is shown in Figure~\ref{fig:flowmodelcorner}, alongside samples from KDE and the training samples.
The trained normalising flow captures the primary features of these distributions, including both the marginal features and the correlations.
Both the KDE and the normalising flow struggle to capture the density of sharp features, such as the height of the spike in $\chi_{\mathrm{eff}}$.
However, the increased flexibility of the normalising flows compared to KDEs mean that this feature is in general better matched by the normalising flow than the KDEs across the training populations, \changed{as shown in the inset plot of Figure~\ref{fig:flowmodelcorner}}. 

One difference observed between the normalising flow and the training samples is in the discrete bands observed at high redshifts. 
These are artefacts of the population synthesis models being generated at discrete lookback times before being combined. 
This binning in lookback time is not an astrophysically expected feature in true populations.
The normalising flow, however, smooths over the binning present in the models in the high redshift tails.
This is a by-product of the normalising flow not having sufficient flexibility (given the network architecture) to pick up on more complicated features, and this feature only being present in the tails of the distribution with few training samples.
This smoothing means that the normalising flows effectively interpolate over simulations at different lookback times at this \ac{GW} observable level.

In order to numerically evaluate the difference between the model representations, we calculate the difference in the KL divergence between the normalising flows and training population models, and the KDE models and the training models. 
For all models across all formation channels, the KL divergence between the normalising flows and the training data is lower than that between the KDE and training data, with an average difference in KL divergence for each formation channel being $\KLdiffCE~\mathrm{nat}$, $\KLdiffCHE~\mathrm{nat}$, $\KLdiffGC~\mathrm{nat}$, $\KLdiffNSC~\mathrm{nat}$, $\KLdiffSMT~\mathrm{nat}$ for the CE, CHE, GC, NSC and SMT channels, respectively. 
\changed{For the difficult $\chi_\mathrm{eff}$ distribution in the CE channel, we find that the Kolmogorov–Smirnov (K–S) test between the normalising flow and the training distribution is lower than the K–S test between the KDE and the training distribution by a factor of $\chieffKSratio$ averaged over all models.
We include the KL and K–S values for individual models in the data release \citep{amazeflowsDR}.}
A lower KL divergence for the normalising flows than the KDEs indicates that the normalising flows are more similar to the training samples than the KDEs.
Therefore, the normalising flows are on average a better representation of the target population synthesis models for each formation channel.

\begin{figure*}
    \centering
    \includegraphics[width=0.72\linewidth]{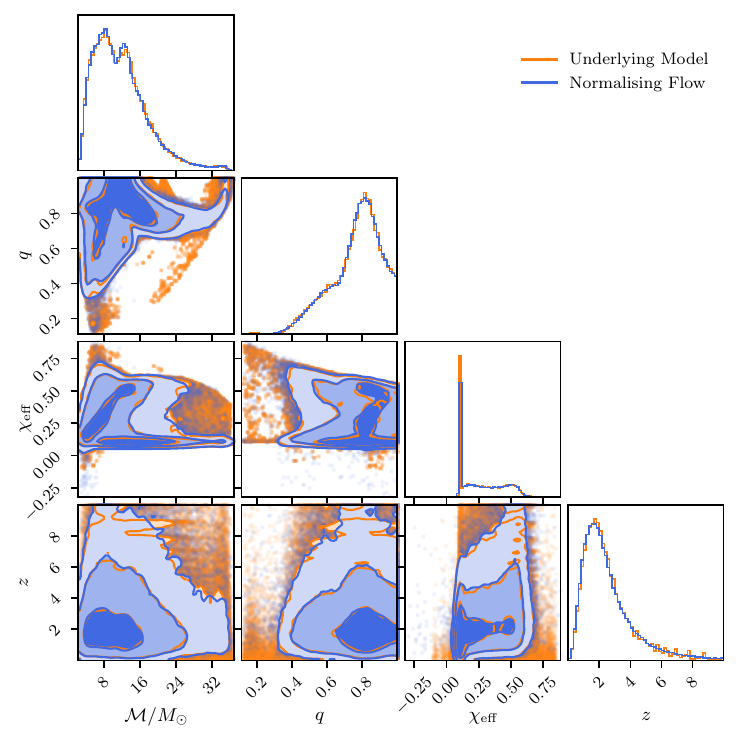}
    \caption{The distribution learned by the normalising flow (blue) for the common envelope channel for $\chib=0.1$ and $\alphaCE=1$, without having seen the underlying training data (orange) for this particular population model. 
    By seeing the training data for other values of $\chib$ and $\alphaCE$ around this population, the normalising flow is able to interpolate and reproduce this unseen distribution.
    The number of effective samples is the same for both the underlying model and normalising flow samples.}
    \label{fig:textmodelcorner}
\end{figure*}

After testing the emulation of the flow at points where it has seen the training data, we also validate our methods by removing one training population from the total training data, and testing how it is reproduced by the normalising flow.
The reproduced unseen population is shown in Figure~\ref{fig:textmodelcorner}.
This model was the $\chib=0.1, \alphaCE=2.0$ training population from the CE channel, a population in the middle of the grid of the $\chib^{\mathrm{*}}$ and $\alphaCE^{\mathrm{*}}$ values used in training.
The remaining $19$ training populations were used to train this normalising flow.
The normalising flow emulates the test population data well even though they were not seen during training. 
\changed{As a figure of merit we compare the KL divergences between the underlying model and the normalising flow with the removed test data, the normalising flow trained on the whole dataset, and the KDE for the population model with $\chib=0.1, \alphaCE=2.0$. 
The KL divergence between the normalising flow at the removed population and the underlying model is comparable to the KL divergence between the normalising flow which saw all the training data and the underlying model, with a difference of $\KLdifftest~\mathrm{nat}$. 
The difference in KL divergences between the normalising flow at the removed population and the underlying model and between the KDE and the underlying model is $\KLdifftestKDE~\mathrm{nat}$. 
There is a smaller difference in KL divergence when comparing the two normalising flows than when comparing the normalising flow with the unseen data to the KDE, meaning the two normalising flows have more similar distributions.
This indicates that the normalising flow trained with the removed population emulates the unseen distribution better than the KDEs, and it has a similar performance to the normalising flows which saw the whole dataset.}
The normalising flows are therefore trained well enough, and with a powerful enough network architecture, to interpolate this set of training data.
Conversely, the interpolation performance shows that there are sufficient training data, and that the training models are close enough together in parameter space to allow for good interpolation.
By showing that the normalising flow can replicate an unseen test distribution, we infer that the normalising flow is reliable across the parameter space where we do not have training samples.

\begin{figure}
    \centering
    \includegraphics[width=0.9\linewidth]{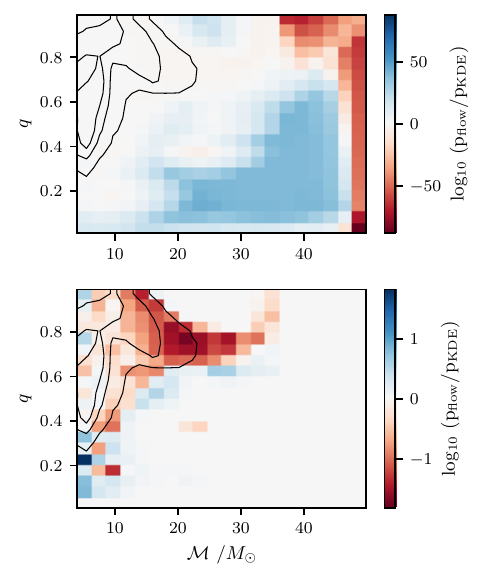}
    \caption{The log likelihood ratio between the normalising flow and the KDE representation of the CE population model with $\chib=0$ and $\alphaCE=5$, at a slice through chirp mass and mass ratio at effective inspiral spin of $0.05$ and redshift of $0.1$. 
    The black contours trace the areas containing $50\%$, $90\%$, and $99\%$ of the training samples in a bin around this slice. 
    The top panel shows no regularisation applied to the normalising flow or the KDE, with the likelihood ratio at the most extreme in the tails of the probability distribution. 
    The bottom panel shows the log likelihood ratio when using regularisation for both representations, minimising the differences in the tails of the distribution. 
    The scale bar in the lower panel covers a much smaller range of ratios.}
    \label{fig:llhratioslice}
\end{figure}

While we have established that KDEs and normalising flows both do well at reproducing the bulk features of the distributions, we find that they do differ in the tails, where we must extrapolate the distributions due to the finite number of samples. 
Our uncertainties here can be ameliorated through our introduction of regularisation (Section~\ref{met:statframe}). 
Figure~\ref{fig:llhratioslice} shows the difference in the likelihood surface $p(\mathbf{d}\mid\vec{\Lambda})$ of the normalising flows in comparison to that of KDEs for a slice through the two-dimensional $\mchirp$--$q$ plane with and without regularisation.
Without regularisation, the biggest difference in the likelihood surface is present in the tails of the distribution where we have few population synthesis samples.
Regularisation mitigates the differences in the tails by adding a uniform value for both the normalising flow and KDE.
Hence, as desired, with the additional regularisation both representations tend to the same (uninformative) limit where we lack simulation points.

\subsection{Testing normalising flows with discrete inference} 
\label{results:discinf}

\begin{figure*}
    \centering
    \includegraphics[width=0.9\linewidth]{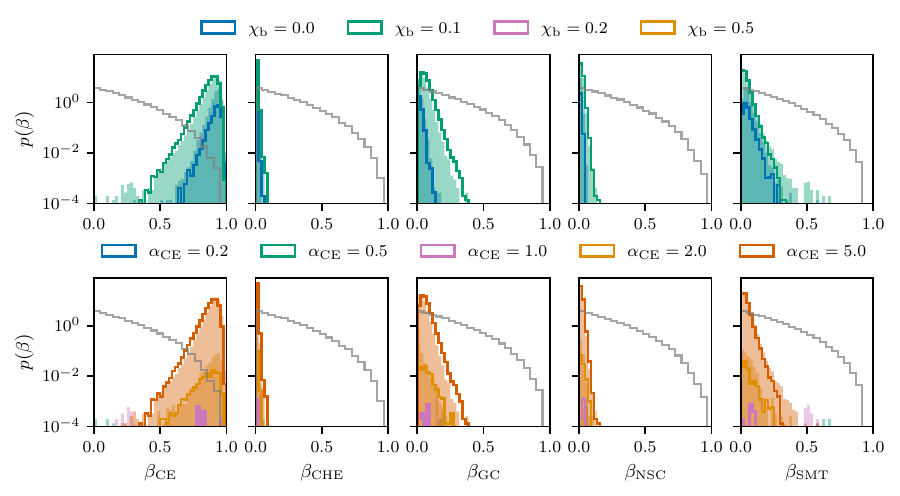}
    \caption{The hyperposteriors on the underlying branching fractions of each formation channel, using the normalising flows (solid lines) and KDEs (filled) for discrete inference. 
    The top row shows the samples corresponding to each of the $\chib$ populations, and the bottom row shows those corresponding to the different $\alphaCE$ models, with the prior in grey. 
    \changed{We infer negligible contribution to the population from the $\chib=0.2$ and $\chib=0.5$ models, or the $\alphaCE=0.2$ model (both with normalising flows and KDE model representations), and hence these curves do not show up in the plot.}}
    \label{fig:discretebranchingfracs}
\end{figure*}

Having validated that the normalising flows can emulate the target population distributions, we now compare the flows to the KDEs in the discrete model selection framework.
After training, we evaluate the probability of the observations given the normalising flows at the points $\{\chib^{\mathrm{*}}, \alphaCE^{\mathrm{*}}\}$.
The hyperposteriors on the underlying branching fractions, and the model selection between the parameters $\chib$ and $\alphaCE$, comparing the two inference methods are shown in Figure~\ref{fig:discretebranchingfracs}, \changed{and posterior samples are included in our data release \citep{amazeflowsDR}}. 
\begin{deluxetable}{l c c}
\tablecaption{
Inferred median and $90\%$ symmetric credible interval for the underlying branching fractions from the discrete inference using normalising flows and using KDEs.
}
\label{tab:discinfbetas}
\tablehead{
 & \colhead{Normalising flows} &  \colhead{KDE}  
}
\startdata
$\beta_{\mathrm{CE}}$  & $\BetaCEdisc^{+\BetaCEdiscup}_{-\BetaCEdisclow}$    & $\BetaCEKDE^{+\BetaCEKDEup}_{-\BetaCEKDElow}$    \\ 
$\beta_{\mathrm{CHE}}$ & $\BetaCHEdisc^{+\BetaCHEdiscup}_{-\BetaCHEdisclow}$ & $\BetaCHEKDE^{+\BetaCHEKDEup}_{-\BetaCHEKDElow}$ \\ 
$\beta_{\mathrm{GC}}$  & $\BetaGCdisc^{+\BetaGCdiscup}_{-\BetaGCdisclow}$    & $\BetaGCKDE^{+\BetaGCKDEup}_{-\BetaGCKDElow}$    \\ 
$\beta_{\mathrm{NSC}}$ & $\BetaNSCdisc^{+\BetaNSCdiscup}_{-\BetaNSCdisclow}$ & $\BetaNSCKDE^{+\BetaNSCKDEup}_{-\BetaNSCKDElow}$ \\ 
$\beta_{\mathrm{SMT}}$ & $\BetaSMTdisc^{+\BetaSMTdiscup}_{-\BetaSMTdisclow}$ & $\BetaSMTKDE^{+\BetaSMTKDEup}_{-\BetaSMTKDElow}$ \\ 
\enddata
\end{deluxetable}

The inferred branching fractions for the five formation channels for both the normalising flows and the KDEs are shown in Table~\ref{tab:discinfbetas}.
The branching fractions are consistent between both methods for all formation channels, with the CE channel having the highest branching fraction.
Both the KDE and the normalising flow prefer the $\chib=0$ and $\chib=0.1$ models, with a slight preference for the $\chib=0.1$ model.
There is a log Bayes factor between the models $\mathrm{log_{10}}\mathcal{B}^{\chib=0.1}_{\chib=0} = \BFchidisc$ for the normalising flows, and $\mathrm{log_{10}}\mathcal{B}^{\chib=0.1}_{\chib=0} = \BFchiKDE$ for the KDEs.
The normalising flow also prefers high $\alphaCE$, in line with the KDE results; both methods strongly preferring the $\alphaCE=5$ model, with some support for $\alphaCE=2$, and minimal support for the models with lower $\alphaCE$.
The log Bayes factor between the two preferred $\alphaCE$ values is $\mathrm{log_{10}}\mathcal{B}^{\alphaCE=5}_{\alphaCE=2} = \BFalphaCEdisc$ for the normalising flows and $\mathrm{log_{10}}\mathcal{B}^{\alphaCE=5}_{\alphaCE=2} = \BFalphaCEKDE$ for the KDEs.
The general agreement of the normalising flows and the KDEs on the hyperposteriors of the branching fractions and the model selection of $\chib$ and $\alphaCE$ indicates the robustness of the results, and shows that normalising flows can perform comparably to the established KDE approach in a hierarchical inference problem.

The branching fraction hyperposteriors for the CE and GC channels are wider for the normalising flow method, indicating that some of the degeneracies between these channels may be driven by the choice of model representation.
The support for natal spin model differs somewhat between the normalising flows and the KDEs, with the KDEs giving a stronger preference for $\chib=0$.
This may be due to the differences in each model representation's ability to capture the sharp spikes in the effective inspiral spin distributions.
The KDEs also have a stronger preference for $\alphaCE=2$.
In general for this inference framework, $\alphaCE$ is less well measured than $\chib$ due to only being constrained by one of the formation channels.
The minor updates to the \ac{AMAZE} framework have not changed the discrete inference results significantly, and results are consistent with past analyses \citep{zevinOneChannelRule2021, Cheng:2023ddt}.
As the model representations are different in some areas of the population parameter space, we expect some minor differences in the inference results, but the overall agreement indicates that the differences are not significant.

\subsection{Continuous inference with GW observations}
\label{results:continf}

\begin{figure*}
    \centering
    \includegraphics[width=0.9\linewidth]{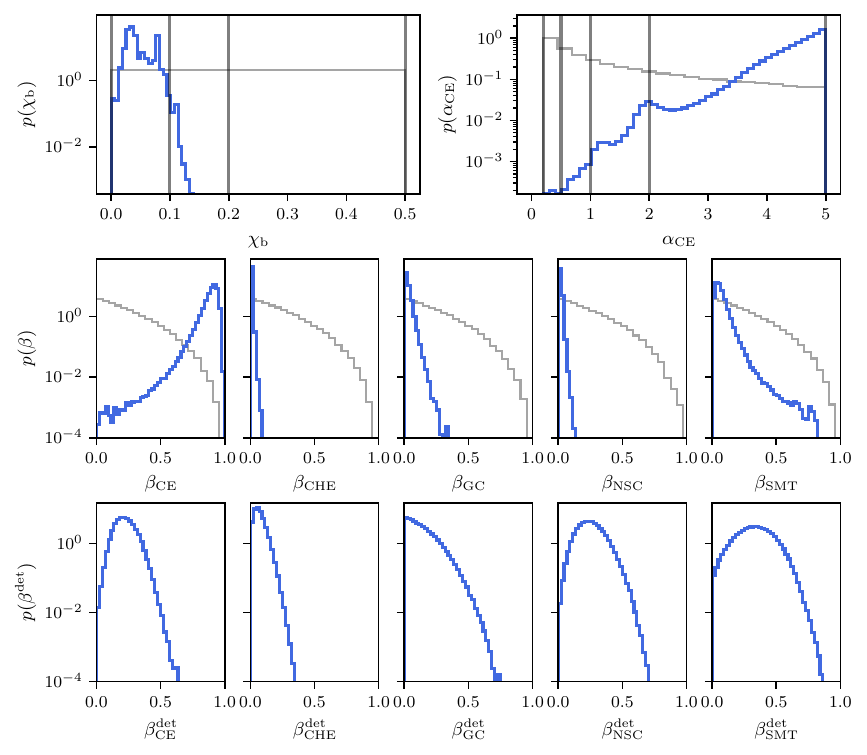}
    \caption{The hyperposteriors on astrophysical hyperparameters $\chib$ (top left) and $\alphaCE$ (top right), the underlying branching fractions for the five formation channels (middle row), and the detectable branching fractions (bottom row), using normalising flows for continuous inference of $\chib$ and $\alphaCE$, and using data from GWTC-3 \citep{LIGOScientific:2021djp}. 
    The vertical lines on the top row show the location for the population synthesis models used to train the normalising flows. 
    The grey histograms show the priors on $\chib$, $\alphaCE$, and the underlying branching fractions.}
    \label{fig:contresult_GWTC3}
\end{figure*}

Having verified the emulation and checked agreement with discrete inference, we now show the unique feature of the normalising flows: the extension to continuous inference.
We allow the astrophysical population parameters $\vec{\lambda}$ to vary anywhere within the prior, inferring a continuous hyperposterior on these parameters.
Our recovered hyperposteriors on $\vec{\lambda}$ and the underlying and detectable branching fractions between formation channels are shown in Figure~\ref{fig:contresult_GWTC3}.
The upper panels show the hyperposteriors (blue histograms) with the points in hyperparameter space where our population synthesis models lie (where the discrete inference was evaluated; grey vertical lines).
The measured values on our hyperposteriors are quoted in Table~\ref{tab:contresults}, \changed{and posterior samples are included in our data release \citep{amazeflowsDR}}.
\begin{deluxetable}{l c c}
\tablecaption{
Inferred median and $90\%$ symmetric credible intervals for the underlying and detectable branching fractions and the natal spin, and the $90\%$ credible lower limit for CE efficiency from the continuous inference results using normalising flows.
}
\label{tab:contresults}
\tablehead{
 & \colhead{$\beta$} &  \colhead{$\beta^{\mathrm{det}}$}  
}
\startdata
CE  & $\BetaCEcontall$  & $\BetaDetCEall$ \\
CHE & $\BetaCHEcontall$ & $\BetaDetCHEall$\\
GC  & $\BetaGCcontall$ & $\BetaDetGCall$ \\
NSC & $\BetaNSCcontall$ & $\BetaDetNSCall$ \\
SMT & $\BetaSMTcontall$ & $\BetaDetSMTall$ \\ 
\midrule
$\chib$           & \multicolumn{2}{c}{$\chibcontall$}    \\
$\alphaCE$         & \multicolumn{2}{c}{$>\alphaCEcontlowlim$}  \\ 
\enddata
\end{deluxetable}

We measure $\chib=\chibcontall$, with most of the posterior support in between $\chib=0$ and $\chib=0.1$.
Given the assumptions in our population models, we find that black holes in merging \acp{BBH} are born with low, but non-zero spin.
We find that $\chib>0$ is preferred over $\chib=0$, with a log Bayes factor $\log_{10}\mathcal{B}_{\chib=0}^{\chib>0}=\chibcontSDR$.

The narrow hyperposterior on $\chib$ may be an artefact of the assumption in the population models that black holes share a common birth spin in the absence of spin-up effects \changed{\citep{zevinOneChannelRule2021}}. 
This assumption causes the sharp feature at $\chib$ in the $\chi_\mathrm{eff}$ distributions  (Figure~\ref{fig:flowmodelcorner}).
\changed{Our models for $\chib=0$ follow the assumption of efficient angular momentum transfer, meaning that} black holes uninfluenced by binary interactions (typically including the first-born black hole for the CE and SMT channels) are born with negligible spin \citep{Fuller:2019sxi}. 
In this case, natal spins are expected to be in a narrow range $\chib \lesssim 0.01$.
The assumption that black holes are born with low spin may not hold when considering different prescriptions for angular momentum transfer or tidal spin up \citep{Belczynski:2017gds,Bavera:2020inc,Olejak:2021iux,Fuller:2022ysb}.
For example, with less efficient angular momentum transfer, we expect the distribution of effective inspiral spin of \acp{BBH} to broaden \citep[e.g.,][]{Perigois:2023ihi}. 
Other influences on the natal spin of a black hole include formation kicks \citep{Baibhav:2024rkn} and accretion during black hole formation \citep{Issa:2025jzq}, while super Eddington accretion could increase the spins of black holes post-formation \citep{vanSon:2020zbk, Bavera:2020uch, Zevin:2022wrw}.
Larger spins are considered as a proxy for these influences within the population models we use by increasing $\chib$; however, the models with $\chib>0$ maintain the narrow range of natal spin values. 
Therefore our use of $\chib$ may not accurately reflect mechanisms that produce higher natal spins. 
Updating the assumptions pertaining to natal spin, angular momentum transfer, and spin up in our population synthesis training data could give us a more reliable \changed{spin measurement}.

Our results for $\alphaCE$ show preference for high values, with $\alphaCE>\alphaCEcontlowlim$ at $90\%$ credibility, and an extended tail towards lower values.
These results motivate follow-up population synthesis simulations at points of interest, at high values of $\alphaCE$ where we have the most support (and at higher values than in our training set), and around $\alphaCE=2$.
These extra simulations could improve the emulation accuracy of the normalising flow around these points.
The populations with highest $\alphaCE$ correspond to populations with the lowest masses and greatest redshifts, hence the lowest detection efficiency.
The detection efficiency increases monotonically from $\alphaCE=1$ to $\alphaCE=5$, therefore the structure we see over this range is not driven by the detection efficiency.

These results reflect the findings from \cite{zevinOneChannelRule2021} and \cite{Cheng:2023ddt} that there is a preference for high $\alphaCE$ when considering the same formation channels and hyperparameters.
This is also in line with \cite{Bouffanais:2020qds}, who found the largest support at $\alphaCE\sim5$ and $\alphaCE\sim6$ when considering only \acp{BBH} from isolated evolution, and $\alphaCE$ in the range $[1,10]$.
\cite{Wong:2020ise} found a weak preference for low $\alphaCE$, using mass and redshift observations to constrain a population made up of a mixture of binaries from isolated evolution and globular clusters. 
However, this was not well constrained as the CE channel was not found to be the majority channel in their results.
The modelling of the CE phase using the $\alphaCE$ prescription may not be best suited to these analyses, as there is growing evidence that $\alphaCE$ is not a universal value for all systems \citep{Politano:2004th, Iaconi_2019}.
Updated prescriptions for the CE phase \citep[e.g.,][]{Hirai:2022ctd} may benefit our constraints on the inference of CE astrophysics.

We find that the measurements of $\alphaCE$ and $\chib$ have no strong correlations with each other or any of the branching fractions.
This shows that despite overlap in the populations, there are sufficiently distinguishing features on a population level given the current observations. 

The branching fraction results are consistent with our discrete inference, showing support for a mix of channels contributing to the observations, with the CE channel contributing the most to the underlying mergers.
We infer the majority of astrophysical mergers to come from the CE channel, despite the prior preference against any single branching ratio being large.
However, the CE channel has the lowest detection efficiency of the formation channels, and thus contributes significantly less to the detected population, as seen in the detectable branching fractions.
The second largest contribution to the underlying channels is the SMT channel, for which we infer a higher branching fraction compared to the discrete inference.
We find a strong correlation between $\beta_{\mathrm{CE}}$ and $\beta_{\mathrm{SMT}}$.
These populations contribute to the lower masses and effective inspiral spins close to $\chib$, and therefore may be harder to distinguish than the other channels with our observations. 

\subsection{Comparing the inferred population to parametric models}
\label{results:dataspace}

\begin{figure*}
    \centering
    \includegraphics[width=0.85\linewidth]{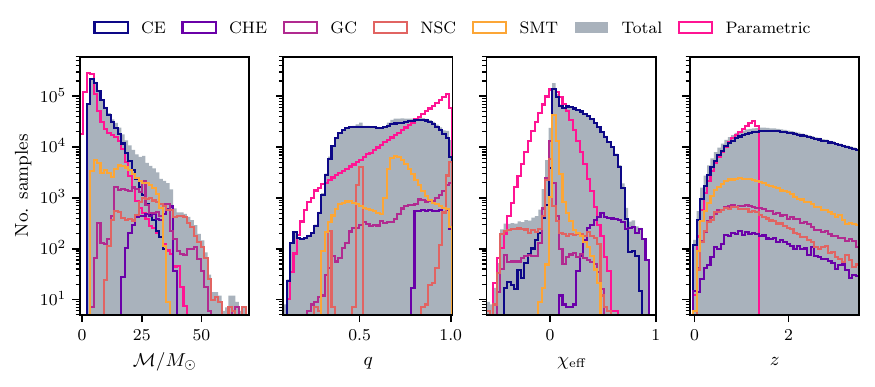}
    \caption{The distribution of GW observables given by the inferred hyperparameters using continuous inference, broken down into contributions from each formation channel, marginalised over the hyperposterior. 
    The total contribution from all five channels is shown in dark grey. 
    We compare this to inferred results using standard parametric models using GWTC-3 \citep{theligoscientificcollaborationPopulationMergingCompact2022}, shown in pink. Compared to the parametric models, our inferred population has a greater contribution at higher chirp masses, mass ratios around $0.5$, and more positive effective inspiral spins.}
    \label{fig:dataspace_result}
\end{figure*}

Finally, we look at the inferred population in \ac{GW} observable-space, given our results from the continuous inference using normalising flows.
We compare these distributions to results from parametric models that are based on simple functional forms (and therefore are less reliant on specific astrophysical assumptions).
We show the parametric inference results from the default models used in the GWTC-3 population analysis \citep{theligoscientificcollaborationPopulationMergingCompact2022}.
This used a power-law-plus-peak model in primary mass \citep{talbotMeasuringBinaryBlack2018}, a power law in mass ratio, an identical beta distribution for both spin magnitudes, a mixture of isotropic and normally distributed spin tilts, and a power law in redshift.
Figure~\ref{fig:dataspace_result} shows the contribution of the five formation channels, marginalised over the hyperposteriors from our continuous inference, in comparison with the parametric inference.

Our measured chirp mass distribution has support out to higher masses in comparison to the parametric model.
While the CE and SMT channels are the primary contributors to the population between $\mathcal{M}\sim 2$--$30~M_{\odot}$, the higher chirp mass tail is from hierarchical mergers in the dynamical formation channels, GC and NSC.
Our results support the notion that the underlying \ac{BBH} population has contributions to the high mass population from dynamical channels \citep{Kimball:2020qyd,Tagawa:2020qll,Antonini:2022vib,Antonini:2024het}.
Both the parametric result and our result suggest that the data supports a large fraction of \acp{BBH} at low masses.
The distribution from the parametric result peaks at smaller masses than our result; this is a consequence of a minimum chirp mass of $\sim2.3~M_\odot$ in the population synthesis models we consider.
The CE channel is the only channel than can contribute to the population for $\mathcal{M} \sim 2.3$--$3.5~M_\odot$, as the other channels do not have chirp masses this low.
This necessitates that the majority of the contribution at low masses comes from the CE channel.
As lower mass \acp{BBH} make up the majority of the underlying population, this drives the underlying branching fraction for the CE channel to be high. 
Exploring a variable minimum black hole birth mass as an input hyperparameter may help to accurately fit the mix of formation channels, and also uncover details of supernova mechanisms \citep{fryerCOMPACTREMNANTMASS2012,Zevin:2020gma,Mandel:2020cig,Farah:2021qom,Olejak:2022zee}. 

There is more structure in our inferred mass ratio distribution than the power-law mass ratio model.
While the power-law model peaks at $q=1$, we find a flatter distribution above $q\sim0.5$, with an excess of mergers in the range $q\sim0.3$--$0.8$ and a dearth of mergers at symmetric mass ratios compared to the power-law model.
Our results are similar to those found by \cite{godfreyCosmicCousinsIdentification2023}, who use a data-driven model to infer the distributions of \acp{BBH} in GWTC-3, finding a flatter distribution of mass ratios between $q\sim0.3$ and $q=1$.
Similarly, \cite{Rinaldi:2023bbd} find a mass ratio distribution which does not show support for equal mass binaries, instead the main contribution being centred around $q\sim0.7$.
The results from \cite{callisterParameterFreeTourBinary} also infer a mass ratio distribution that either peaks at $q=1$ or is flat in mass ratio.
While the parametric power-law model is limited to a monotonic function of mass ratio, these data-driven models identify alternative structure in the data.
This raises the question whether \acp{BBH} do preferentially merge with equal masses \citep{Fishbach:2019bbm}, or if this is driven by assumptions in the power-law model.
Our results add to the evidence that the mass ratio distribution of \acp{BBH} does not peak at $q=1$, motivating the investigation of alternative parametric models for this distribution. 

For the effective inspiral spin distribution, our inferred population is significantly more asymmetric than the parametric distribution, with a positive skew. 
Our population models assume that \acp{BBH} with large negative effective inspiral spin can only be produced by the two dynamical channels, and we infer only a small fraction of mergers come from dynamical channels.
Furthermore, our constraints on $\chib$ to low values mean that there is a narrower spread of $\chi_\mathrm{eff}$ for the dynamical channels, contributing to the lack of support for negative $\chi_{\mathrm{eff}}$ in our measurements.
There is a general preference for positive $\chi_\mathrm{eff}$ in the data \citep[e.g.,][]{Miller:2020zox, Roulet:2021hcu, callisterWhoOrderedThat2021, theligoscientificcollaborationPopulationMergingCompact2022, Edelman:2022ydv, Adamcewicz:2023szp}.
Recent work by \cite{Banagiri:2025dxo} find a positively skewed $\chi_\mathrm{eff}$ distribution, although the distribution is mostly contained to $\chi_\mathrm{eff}<0.4$.
In comparison, we find the fraction of the population with $\chi_\mathrm{eff}>0.4$  to be $\highchibfrac\%$, compared to \changed{$\skewhighchibfracparam\%$ from the \texttt{skewnormal} mixture model from \cite{Banagiri:2025dxo}, and} $\highchibfracparam\%$ for the parametric results from the default models used in the GWTC-3 population analysis \changed{\citep{theligoscientificcollaborationPopulationMergingCompact2022}}.
Our results show a larger preference for positive $\chi_\mathrm{eff}$, in contrast to these previous works. 
\changed{The \texttt{skewnormal} mixture model from \cite{Banagiri:2025dxo} allows for a subpopulation with high $\chi_{\mathrm{eff}}$, but the data do not strongly support such a feature. 
This suggests that the significant contribution from \acp{BBH} with $\chi_\mathrm{eff}>0.4$ that we infer could be driven by the strong astrophysical assumptions in our population models. 
However, our results are in agreement with these other works in that there is negligible support for a highly spinning subpopulation of \acp{BBH}, with our results finding that $99\%$ of the inferred population has $\chi_{\mathrm{eff}}<\upperchieffpercentile$.
These comparisons between population models highlight the usefulness of models with different strengths of astrophysical assumptions in understanding what drives the shape of our inferred population in different regions of parameter space.
}

The redshift distribution varies the least between formation channels, and looks the most similar to the parametric result.
The parametric distribution is cut off at $z=1.35$ as our \ac{GW} measurements are not strongly constrained beyond this range.
Our result has a shallower redshift evolution than the parametric model.
With measurements of \acp{GW} past $z>2$, measuring the peak of the redshift distribution and the rate evolution at higher redshifts could help distinguish between different formation channels.

\section{Conclusion} \label{sec:conc}

We extended the \ac{AMAZE} framework to continuous population inference over astrophysical hyperparameters, using normalising flows to interpolate simulations from population synthesis.
We trained normalising flows to emulate the distributions of GW observables for five different formation channels, given two astrophysical input hyperparameters.
Through this method we demonstrated how we can use astrophysically motivated population models of \ac{BBH} populations to understand the origins of our observed population, without the computational expense of a dense grid of population synthesis simulations. 

We validated our method with comparisons to a non-interpolating method which uses KDEs for the astrophysical model representation, ensuring that the normalising flows are fit for purpose.
These tests show that the flexibility of the normalising flows can capture features in our training populations more accurately than the KDEs.
Both methods produce consistent results when used for inference between the discrete training population models. 

Using GWTC-3 observations \citep{LIGOScientific:2021djp}, we applied the normalising flow interpolation to recover continuous hyperposteriors on our astrophysical hyperparameters (natal spin and CE efficiency) alongside the branching fractions of the formation channels. 
Given our population model, the measurement of natal spin suggests that black holes are born with small spins of $\chib=\chibcontall$.
We find a preference for high CE efficiency, with $\alphaCE>\alphaCEcontlowlim$ at $90\%$ credibility.
This pushes to the edge of our prior on $\alphaCE$, motivating further population synthesis simulations around and above $\alphaCE=5$. 
While detailed simulations of CE evolution have indicated that $\alphaCE$ may be greater than $1$, it is not yet clear what the largest physically plausible description may be for \ac{BBH} formation \citep{Fragos:2019box,Law-Smith:2020jwf,Lau:2021jpm,Roepke:2022icg}.
In general, our framework can be used to find the region of highest (and lowest) probability for population synthesis input hyperparameters, and determine interesting locations for further simulations. 

Our analysis indicates that the majority of mergers (in the underlying population) come from the CE channel, in line with our tests with discrete inference and the results from \cite{Cheng:2023ddt}.
Due to the much lower detection efficiency for this channel, we find that the detected fraction of CE binaries is $\BetaDetPercCEall\%$ of the overall population, and no one channel dominates the observed population. 
Given this mix, it is important not to interpret all observations as coming from a single channel. 

The continuous inference framework is crucial for future investigations of key aspects of \ac{BBH} formation. 
Measurements of astrophysical hyperparameters vary when making different assumptions about the formation channels present \citep{zevinOneChannelRule2021, Cheng:2023ddt}, and their possible input parameters.
For example, we measure higher CE efficiency than \cite{Wong:2020ise}, partially driven by the choice of population inputs.
Considering more formation channels in our analysis means that a single channel does not have to account for a wide range of observations.
This motivates the inclusion of multiple channels and astrophysical hyperparameters in future analyses.
Our framework employing normalising flows for continuous inference is well suited to population inference with multiple formation channels and varying astrophysical hyperparameters.

We find that our inferred population shows more structure in the distributions of chirp mass, mass ratio and effective inspiral spin than current results using parametric population models \citep{theligoscientificcollaborationPopulationMergingCompact2022}. 
Using population synthesis models provides a more astrophysically informed approach to interpreting observations. 
There are many potential uncertainties in these models, which may influence our conclusions. 
Using our continuous inference, it is possible to marginalise over the uncertainty in input hyperparameters, mitigating potential bias from uncertainty in these parameters. 
However, our results are only as reliable as the models used as input. 
Furthermore, we do not yet account for the absolute merger rate predicted by our population models, instead marginalising over this quantity.
This means that the predicted rate from the population synthesis may not match the observed rate of \ac{BBH} mergers; additionally, adding in information on the absolute merger rate could make results more precise \citep{barrettAccuracyInferencePhysics2018,Mastrogiovanni:2022ykr}.
By comparing our results to parametric and data-driven population models, we may further check the consistency of results, and explore the limitations of population models.
Our results illustrate the \changed{complementary} nature of the different kinds of population analyses, and motivate including new structure to parametric models.

While we show that the normalising flows can effectively interpolate the population distributions over our hyperparameter space, for the detection efficiency we used a cubic polynomial interpolation. 
This encodes the assumption that the detection efficiency values we calculate from the population synthesis models are representative for the full model space we use in the continuous inference, which may not necessarily be true.
The accuracy of this quantity is especially important with higher numbers of \ac{GW} observations, as it scales to the power of the number of observations.
Previous work has shown that neural networks can be used to emulate the selection function for \ac{GW} detection \citep[e.g.,][]{mouldDeepLearningBayesian2022, Chapman-Bird:2022tvu, Callister:2024qyq}.
A similar approach could be implemented in our framework, extending our implementation of neural networks to interpolate the detection efficiency, making our results more reliable.

More \ac{BBH} observations will give us more constraining power to measure \ac{BBH} population properties \citep{Zevin:2017evb, barrettAccuracyInferencePhysics2018}.
We expect hundreds more observations from the LIGO--Virgo--KAGRA fourth observing run \citep{KAGRA:2013rdx}.
These will allow us to measure a larger subset of input hyperparameters to population synthesis.
To fully leverage our population models with these observations, it is necessary that the models reflect our uncertainty on the astrophysical origins of \acp{BBH} and consider these larger sets of parameters.
Our work illustrates that using normalising flows for astrophysical inference is a robust method that is capable of continuous inference over multiple population hyperparameters and multiple formation channels, with potential to be extended to higher dimensional inference.
In the future, this method may be extended to infer additional parameters such as black hole natal kick magnitude, stellar cluster properties, and stellar wind parameters.
With more population synthesis simulations varying these parameters, our continuous inference framework could be used to understand the degeneracies between these key variables in \ac{BBH} progenitor evolution, and uncover the origins of these systems. 

\section*{Acknowledgments}
We thank Christian Chapman-Bird, Jessica Irwin, Narenraju Nagarajan, Federico Stachurski, and Michael Williams for useful discussions, Matthew Mould \changed{and the anonymous referee} for insightful comments on the manuscript, and the groups at Caltech and Northwestern for hospitality while this work was developed.  
SC is supported by \ac{STFC} studentship 2748218.
CPLB and JV are supported by \ac{STFC} grant ST/V005634/1. 
MZ gratefully acknowledges funding from the Brinson Foundation in support of astrophysics research at the Adler Planetarium.
SC thanks the \ac{CIERA} Board of Visitors and the University of Glasgow MacRobertson Scholarship for support to visit \ac{CIERA} to complete part of this study, and thanks \ac{CIERA} for their hospitality. 
This research has made use of data from the Gravitational Wave Open Science Center, a service of the LIGO Scientific Collaboration, the Virgo Collaboration, and KAGRA. This material is based upon work supported by NSF's LIGO Laboratory which is a major facility fully funded by the National Science Foundation, as well as \ac{STFC} of the United Kingdom, the Max-Planck-Society (MPS), and the State of Niedersachsen/Germany for support of the construction of Advanced LIGO and construction and operation of the GEO\,600 detector. Additional support for Advanced LIGO was provided by the Australian Research Council. 
Virgo is funded, through the European Gravitational Observatory (EGO), by the French Centre National de Recherche Scientifique (CNRS), the Italian Istituto Nazionale di Fisica Nucleare (INFN) and the Dutch Nikhef, with contributions by institutions from Belgium, Germany, Greece, Hungary, Ireland, Japan, Monaco, Poland, Portugal, Spain. KAGRA is supported by Ministry of Education, Culture, Sports, Science and Technology (MEXT), Japan Society for the Promotion of Science (JSPS) in Japan; National Research Foundation (NRF) and Ministry of Science and ICT (MSIT) in Korea; Academia Sinica (AS) and National Science and Technology Council (NSTC) in Taiwan. 
This document has been assigned LIGO document number \href{https://dcc.ligo.org/LIGO-P2500074/public}{LIGO-P2500074}.

These results were produced using the updated \ac{AMAZE} framework, found in \href{https://github.com/michaelzevin/AMAZE/tree/v2.0.0}{this code release}. 
Results and data products used in this work are available on Zenodo: \dataset[doi:10.5281/zenodo.14967687]{https://doi.org/10.5281/zenodo.14967687} \citep{amazeflowsDR}.

\software{\ac{AMAZE} \citep{zevinOneChannelRule2021}, \texttt{glasflow} \citep{glasflow}, \texttt{nflows} \citep{nflows}, weights and biases \citep{wandb}, \texttt{emcee} \citep{Foreman_Mackey_2013}, \texttt{NumPy} \citep{harris2020array}, \texttt{SciPy} \citep{2020SciPy-NMeth}, \texttt{pandas} \citep{reback2020pandas}, \texttt{matplotlib} \citep{Hunter:2007}, \texttt{corner} \citep{corner}}

\appendix

\section{Training Normalising Flows}
\label{apx:trainingflows}

We train one normalising flow for each of the five formation channels. 
We wish to train the normalising flows to reconstruct the population distribution $p_{\mathrm{pop}}(\vec{\theta}\mid\vec{\lambda})$ using samples of parameters $\vec{\theta}$ from population synthesis simulations. 
To improve the performance of the training, we use scaled and transformed versions of the source parameters, rather than using $\vec{\theta}$ directly. 
First, we use logistic or hyperbolic mappings such that each dimension in the observable data space $\theta$ is in the range ($-\infty$,$\infty$): 
\begin{equation}
    \begin{split}
        \tilde{\mathcal{M}} = \logit\left(\frac{\mathcal{M}}{\mathcal{M}^\mathrm{up}}\right),\\
        \tilde{q} = \logit\left(\frac{q}{q^\mathrm{up}}\right), \\
        \tilde{\chi}_\mathrm{eff} = \arctanh(\chi_\mathrm{eff}),\\
        \tilde{z} = \logit\left(\frac{z}{z^\mathrm{up}}\right), 
    \end{split}
\end{equation}
where the inverse logistic function on $p$ is defined as
\begin{equation}
    \logit(p) = \log\left(\frac{p}{1-p}\right), \, 0<p<1, 
\end{equation}
and $\{\mathcal{M}^\mathrm{up}, q^\mathrm{up}, z^\mathrm{up}\}$ are convenient upper bounds.  
The upper bounds are set to $\mathcal{M}^\mathrm{up} = 100 M_{\odot}$ $q^\mathrm{up} = 1$ and $z^\mathrm{up}= 10$, except for the GC channel, which contains samples where $q=1$ and therefore $q^\mathrm{up} = 1.001$ to ensure $0<q/q^\mathrm{up}<1$. 
Finally, the transformed parameters are rescaled to give 
\begin{equation}
    \label{theta'mappings}
    \begin{split}
        \vec{\theta '} = &\ \left\{\frac{\tilde{\mathcal{M}}}{\max(\tilde{\mathcal{M}})}, \frac{\tilde{q}}{\max(\tilde{q})}, \tilde{\chi}_\mathrm{eff}, \frac{\tilde{z}}{\max(\tilde{z})}\right\}.
    \end{split}
\end{equation}
This rescaling is empirically chosen such that $\vec{\theta '}$ is defined on a similar domain across each dimension.
The normalising flows are trained on samples of $\vec{\theta'}$ conditional on $\vec{\lambda}$, learning the distributions $p_\mathrm{pop}(\vec{\theta'}|\vec{\lambda}, \mathrm{channel})$.

Before training, $20\%$ of the population synthesis samples were randomly chosen across different values of $\vec{\lambda}$ as the validation set used in the training.
The normalising flows were implemented using CouplingNF spline normalising flows \citep{durkanNeuralSplineFlows2019} using the \texttt{glasflow} \citep{glasflow} package. 

The specific parameters of the normalising flows are shown in Table~\ref{tab:flowhyperparams}. 
These parameters were chosen considering the results of network parameter optimisation tool Weights and Biases \citep{wandb}. 
\changed{We used Weights and Biases to train 20 normalising flows for each formation channel, varying the number of transforms, neurons, and spline bins for each instance of training. 
We considered number of transforms in the range $[4,10]$, number of neurons in the range $[10,130]$, and number of spline bins in the range $[3,8]$. 
Bayesian optimisation was used to find the network parameters that best minimised the validation loss \citep{BayesianOpt2016}. 
For the NSC and SMT channels, the choice of network parameters did not make a significant difference on the minimum validation loss. 
For the other channels, normalising flows with higher numbers of neurons resulted in lower validation losses. 
We chose the number of transforms and spline bins (given in Table~\ref{tab:flowhyperparams}) to reduce the computational expense while still retaining the flexibility to emulate the training populations.
When visually inspecting the emulated populations, we found that the CE and NSC channels better captured the target correlations with $5$ spline bins than with $4$ spline bins. 
The number of training epochs was chosen such that the validation loss had plateaued by the end of training when employing the chosen network configuration; our training and validation-loss data are included in the data release \citep{amazeflowsDR}.}
In addition, the learning rate was annealed with cosine annealing throughout training \citep{loshchilov2017sgdrstochasticgradientdescent}, with an initial learning rate of $0.001$. 
The training was performed on using a NVIDIA GeForce RTX 2080 Ti GPU and took $\sim50~\mathrm{hr}$ for five channels.
\begin{deluxetable}{l c c c c c}
\tablecaption{
The network parameters of the normalised flows, including the number of training epochs, number of transformations, number of neurons per layer, and number of spline bins in each transform. 
These parameters vary for the five normalising flows, which each learn the populations in a given formation channel.
}
\label{tab:flowhyperparams}
\tablehead{
\colhead{Formation channel} &  \colhead{Epochs} & \colhead{Transforms} & \colhead{Neurons} & \colhead{Blocks} & \colhead{Spline bins} 
}
\startdata
CE                  & $15000$ & $6$  & $128$ & $2$ & $5$   \\
NSC                 & $10000$ & $6$  & $128$ & $2$ & $\changed{5}$   \\
CHE, GC, SMT        & $10000$ & $6$  & $128$ & $2$ & $\changed{4}$   \\
\enddata
\end{deluxetable}

\bibliography{refs}{}
\bibliographystyle{aasjournal}

\end{document}